 \documentclass[twocol]{ametsocV6.1}

\title{Perturbing the Surface Energy Balance to Emulate the Historical Pattern of Tropical Pacific Sea Surface Temperature Trends}

\authors{Timothy M. Merlis,\aff{a}\correspondingauthor{Timothy M. Merlis, tmerlis@princeton.edu}}

\affiliation{\aff{a}{Program in Atmospheric and Oceanic Sciences, Princeton University}}

\abstract{ The strengthening of the zonal sea surface temperature
  (SST) gradient observed in the tropical Pacific in recent decades is
  a regional climate change signal that may be outside the range of
  historical simulations with comprehensive climate models.  Given the
  important role that this change has on other aspects of climate, a
  series of idealized surface energy balance calculations with imposed
  parameters is performed to build a baseline understanding of the
  sensitivities that govern these changes.  I quantify the requisite
  magnitudes of five perturbations that reach a new equilibrium with a
  mean-SST warming of about $0.5 \, \mathrm{K}$ and about $0.4 \,
  \mathrm{K}$ more west Pacific warming than east Pacific warming,
  based approximately on observed trends.
  A characteristic magnitude of zonal asymmetry in a surface energy
  tendency that can bring changes in line with observed trends is
  $\approx 3 \, \mathrm{W \, m^{-2}}$.
  Strengthened zonal SST
  gradients can arise from a more zonally asymmetric ocean heat flux
  that increases by $\approx 20\% \, \mathrm{K}^{-1}$ using that implied by ERA5's
  surface fluxes, a spatially
  varying radiative forcing with a west--east contrast of $\approx 3.3
  \, \mathrm{W \, m^{-2}}$, a more amplifying surface radiative
  feedback in the west than the east with a contrast of $\approx 4 \,
  \mathrm{W \, m^{-2} \, K^{-1}}$, a surface-air relative humidity
  (RH) contrast that increases RH in the west and decreases it in the east
  by $\approx 0.5 \% \, \mathrm{K^{-1}}$, or a more zonally asymmetric wind speed
  that increases by $\approx 16 \% \,
  \mathrm{K^{-1}}$.  The ``storylines'' of forced surface energy
  budget change identified here are valuable in determining the
  plausibility of mechanisms that may be absent or underestimated in
  coupled climate model simulations.
}

\begin{document}

\maketitle

%
%
%
\statement What is the magnitude of changes in energy fluxes in the tropical Pacific
that could provoke the recent several decades of observed sea surface temperature trends?
This urgent question of climate science is addressed using
surface energy balance calculations with the various input terms perturbed
in an idealized manner. These calculations provide an essential order-of-magnitude
baseline for evaluating how novel mechanisms that are absent or weak in comprehensive
climate model simulations might affect the pattern of historical trends.

%
%
%

%

\section{Introduction}

The pattern of sea surface temperature (SST) change is a long-standing
focus area of climate science. The extent to which the tropical
Pacific has mean-state warming that is a reduction of zonal
temperature gradients (i.e., enhanced warming in climatologically cold
east Pacific and resembling the El Ni\~no phase of variability) or
enhancement of zonal temperature gradients (i.e., enhanced warming in
the climatological warm west Pacific) under anthropogenic radiative forcing has been
extensively investigated in climate models \citep[e.g.,][among many
  others]{knutson95,seager19,wills22,seager22}. The observational record has
century-scale trends toward enhanced zonal temperature gradients
\citep{lee22} and the recent four decades are, with a few exceptions,
outside the range of historical simulations in comprehensive coupled
climate model simulations \citep[][their Fig.~2a]{wills22}.

The prospect that the forced response pattern of coupled climate
models is outside the range of observed trends has garnered
substantial interest.
The attention is motivated by the impacts of the
response pattern on the global-mean top of atmosphere radiative flux
\citep[e.g.,][]{andrews22} and regional hydroclimate changes and
extremes \citep{vecchi08b,xie10,sobel23,zhao24}.

One of the primary tools to investigate the forced pattern of
temperature change, comprehensive coupled climate models participating
in various phases of the Coupled Model Intercomparison Project (CMIP), faces
the conceptual challenge of having multiple interacting components to disentangle.
For example, changes in both cloud radiative properties and ocean
circulation can affect the SST pattern individually and they also interact, as is known
from climate sensitivity research \citep[e.g.,][]{winton10,trossman16}.
As such, it
is worthwhile to examine influences on the SST pattern in a manner where aspects of
the climate system can be individually controlled. Here, I pursue
this approach using surface energy balance calculations with key
components prescribed or interactive in a simple
temperature-dependent way. Building the sensitivities of the zonal SST contrast
to external changes or internal temperature-dependent factors
in this
simpler framework is potentially valuable in focusing attention on the
more sensitive components or on more uncertain components in
comprehensive models and establishes a baseline for the requisite
magnitude of proposed `missing' or weak mechanisms in comprehensive
climate models.

Several mechanisms that underlie the radiatively forced pattern of
tropical Pacific SST changes have been described.
\citet{knutson95} 
analyzed the
surface energy budget in a coupled general circulation model (GCM).
They identified ``evaporative damping'' as
the underlying reason for their simulations' weakened gradients: the bulk aerodynamic formula
for evaporation depends on the saturation specific humidity, a temperature-dependent quantity.
At warmer climatological SSTs,
a smaller SST warming can provide the increase in the saturation specific humidity
to rebalance the surface energy budget via evaporative cooling, which tends to reduce
the east--west SST contrast in warming scenarios.
\citet{merlis11}
elaborated this into a scaling theory that captured the zonal SST
contrast in a wide range of idealized atmospheric GCM simulations with
prescribed and climate-invariant ocean heat flux convergence in a
slab ocean, and this mechanism has been diagnosed in comprehensive GCM
surface energy budgets \citep{xie10,zhang14b}. The net surface heat flux contribution by
the ocean circulation is, of course, free to evolve under radiative
forcing. The ocean dynamical thermostat
suggests a strengthening of the zonal ocean heat flux contrast from 
the upwelling of pre-industrial water in the east Pacific \citep{clement96}.
This was initially identified in 
a coupled model of intermediate complexity \citep{clement96}, but
similar patterns can appear transiently in abruptly forced coupled GCMs \citep[e.g.,][]{held10,heede21}.
Further,
\citet{seager19} suggested that biases in GCMs' climatological
relative humidity distribution played a substantial role in their
excessive east Pacific warming. Beyond ocean circulation and the
turbulent surface fluxes' temperature sensitivity, there are radiative
changes that can alter the zonal energy contrast. \citet{dinezio09}
showed a cloud-induced decrease in net shortwave radiation at the surface in the
west Pacific and little change in the east Pacific in the
ensemble-mean of doubled carbon dioxide concentration simulations in
third phase of CMIP. There is also a role for meridional
changes in cloud albedo to affect the zonal SST gradient via shallow
ocean circulations \citep{burls14}. In short, there is a
moist thermodynamic reason to expect reduction in gradients, which
will be modulated by ocean heat transport (and/or differential changes
in storage) and cloud feedbacks. There are additional proposed forced
or feedback-related pathways in addition to those described here
\citep{watanabe24}, all of which would have a surface energy balance
manifestation.

Here, surface energy budget calculations are performed with
prescribed ocean heat transport, radiative forcing, radiative
feedbacks, surface relative humidity, and wind speed to determine how
large changes in these components would have to be to account for a strengthening
of the zonal temperature gradients comparable to the trends in recent
decades.  This is a quantitative framework, but it is not a closed
model of the climate because of the various factors that are imposed.
While the individual perturbations are separately imposed here, they
may vary in concert in Earth's climate; however, they approximately obey linear
superposition, so it is meaningful and simplest to present them
in isolation.
Consideration of the surface budget, rather than the top-of-atmosphere budget,
is valuable in that the ocean heat flux has a direct influence on the SST
and it has been implicated in mechanisms that determine the tropical
Pacific SST change (additional discussion of this choice can be found
in section~\ref{sec-conclusion}).
An analogous
approach has been used for the meridional pattern of temperature
change, where diffusive moist energy balance models with imposed
surface fluxes, radiative forcing and feedbacks have been used to
emulate GCMs \citep[e.g.,][]{roe15,hill22}.
The perturbation surface energy balance has been used as a diagnostic
framework for understanding GCM-simulated changes in the tropical
Pacific SST under external forcing \citep[e.g.,][]{xie10,zhang14b,hwang17,kang23}. 
The goal here, in
contrast, is not to emulate GCMs but to engage with observed changes
of the last four decades.

The analysis presented here is a form of ``storylines'': I aim to answer
what would bring changes in the surface energy balance---a balance
that Earth's climate and all climate models obey---in line with
observed temperature trends.  Conceivably, all of these storylines for
forced responses could be refuted by physical reasoning or
observations. In this case, one would conclude that internal
variability is critical to observed trends. One could pursue the
formalized Bayesian inference that has been used for global climate
sensitivity \citep{stevens16,sherwood20}---this is an
interesting avenue for follow-up for this and other regional aspects of possible comprehensive
model-observed trend discrepancies. 

The paper proceeds as follows. In the next section, I present the surface
energy budget of the tropical Pacific from a reanalysis product and
identify the target SST trend. The approximate surface energy balance
calculations with imposed parameters that are perturbed are also
described in section~\ref{sec-surf_en_budget}.  These
components are individually perturbed to identify parameter values
that bring the SST changes to ones with comparable east-west contrasts
to observed trends---and concomitant changes in the surface
energy balance---are examined in section~\ref{sec-ind_pert}.
Combined or simultaneous perturbations are presented in
section~\ref{sec-comb_pert}.
Last, I discuss the results and conclude.

\section{Surface Energy Budget and SST Trends} \label{sec-surf_en_budget}

\subsection{ERA5 SST and Trends from 1979 to 2023}

Figure~\ref{fig-era_summary}a shows the annual-mean climatology of tropical
Pacific SST and Figure~\ref{fig-era_summary}c shows
the near-equator average Pacific SST (meridionally averaged
within $5^\circ$ of the equator with land regions masked) in the
European Centre for Medium-Range Weather Forecasting (ECMWF)'s ERA5
reanalysis \citep{hersbach20} for the period from 1979 to 2023.
The SST is warmest in the west and decreases by
about $4 \, \mathrm{K}$ by the eastern boundary of the basin, where
there is a shallow thermocline and upwelling of cold water
(Fig.~\ref{fig-era_summary}a,c).

Figure~\ref{fig-era_summary}b,d shows the linear trend of SST:
there is a trend toward enhanced warming in the west and suppressed warming in
the east Pacific. 
The linear
trends imply about $0.6 \, \mathrm{K}$ more warming in the west than the east
for these $44$-year trends, using the near-equator regions of \citet{wills22}
(longitude $110^\circ$ to $180^\circ$ E for the west Pacific and
$180^\circ$ to $100^\circ$ W for the east Pacific). This is somewhat larger than
the trends in the three SST datasets presented in \citet{wills22}, which ended in 2020.
So, a more lax target of $0.4 \, \mathrm{K}$ west-minus-east SST contrast
is set for the perturbation surface energy
balance calculations presented in the next section.
The surface energy balance,
described next, is the lens through which changes in SST will be
examined.

\subsection{ERA5 Surface Energy Budget}

\begin{figure*}[t]
  \noindent\includegraphics[width=32pc,angle=0]{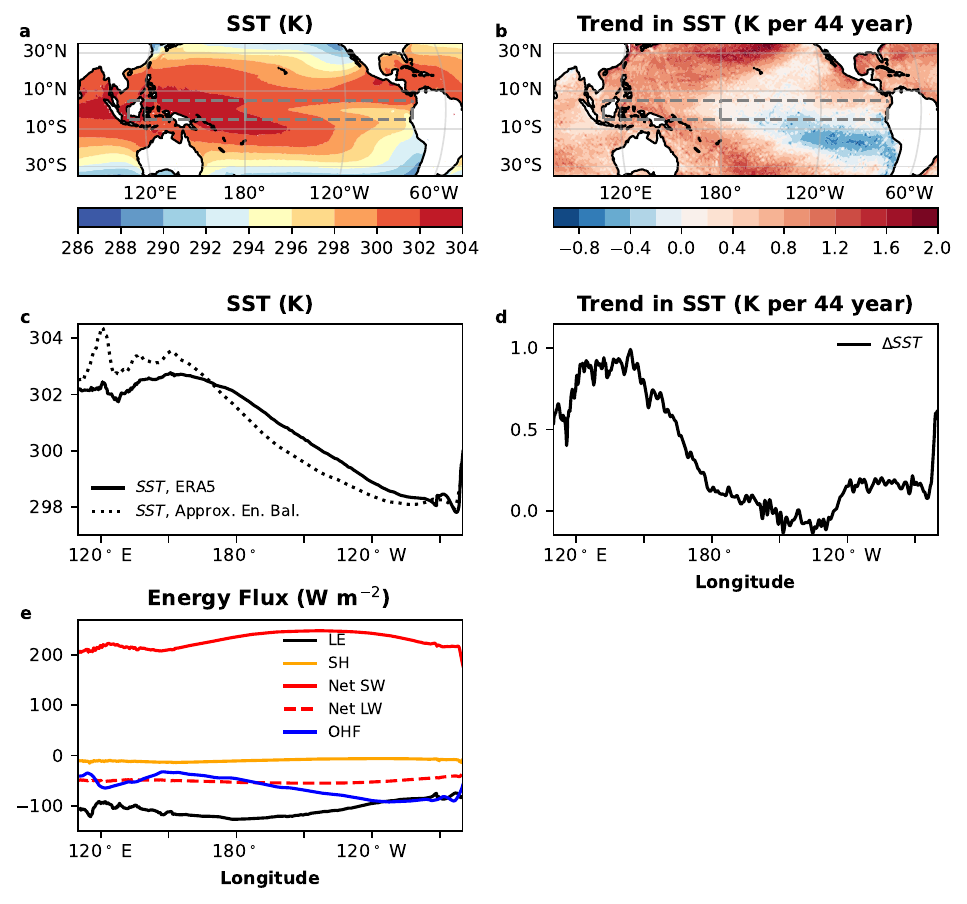}\\
  \caption{(a) Climatological sea surface temperature (SST, units of
    $\mathrm{K}$) of the tropical Pacific and (b) linear trends in SST
    (units of $\mathrm{K}$ per $44$ years) for the same region with
    gray boxed region showing the west and east Pacific averaging
    regions.  (c) Near-equator ($\pm 5^\circ$) average Pacific SST
    with the equilibrium SST from the control approximate surface
    energy balance calculation shown in the dotted black line and (d)
    its linear trend. (e) Surface energy budget tendencies for the
    near-equator average in the Pacific ocean (individual components
    indicated in the legend, units of $\mathrm{W \, m^{-2}}$). All
    quantities are from ERA5 over the period 1979-2023.  Note that the
    longitudinal extent of the plots differs in the top row.
  }\label{fig-era_summary}
\end{figure*}

The surface energy balance is
\begin{equation}
  C \partial_t SST  = R_{SFC} - LE - SH - \mathcal{O} + \mathcal{F},
  \label{eqn-sfc_en}
\end{equation}
with heat capacity $C$, sea surface temperature $SST$, net surface
radiation $R_{SFC}$, latent energy flux $LE$, sensible heat flux $SH$,
ocean heat flux $\mathcal{O}$, and surface radiative
forcing $\mathcal{F}$. The surface radiative forcing $\mathcal{F}$ here includes both
shortwave and longwave radiation perturbations from external forcing,
while the net surface radiation $R_{SFC}$ represents both the climatological
and temperature-dependent perturbation radiative fluxes.
For the ocean mixed layer, the ocean heat flux $\mathcal{O}$ would include
the horizontal ocean heat flux convergence and the vertical ocean heat fluxes
into or out of this layer, both of which can include turbulent or eddy components
of the flow.
The dominant positive energy tendency
of the surface is the net shortwave radiation of $\approx 230 \,
\mathrm{W \, m^{-2}}$ (red line in Fig.~\ref{fig-era_summary}e), and
the net longwave radiation is a negative tendency of the surface of
$\approx -50 \, \mathrm{W \, m^{-2}}$ (red dashed in
Fig.~\ref{fig-era_summary}e).  The dominant negative energy tendency
of the surface is the latent energy flux of $\approx - 110 \, \mathrm{W
  \, m^{-2}}$ (black line in Fig.~\ref{fig-era_summary}e). And the
turbulent sensible heat flux is a modest negative tendency of $\approx
- 10 \, \mathrm{W \, m^{-2}}$ (orange line in
Fig.~\ref{fig-era_summary}e). Assuming a steady state, the residual of
the terms from ERA5 is the implied ocean heat flux 
$\mathcal{O}$ in \eqref{eqn-sfc_en}, which is referred to as
ocean heat flux (OHF) for brevity.  This is a negative tendency in the
deep tropics of the Pacific basin of $\approx -60 \, \mathrm{W \,
  m^{-2}}$ 
\citep[blue in Fig.~\ref{fig-era_summary}e, see also][]{trenberth01b}. 
The departures from the basin mean are important for the pattern of
warming and they are discussed next. 

In this estimate of the Pacific basin's surface
energy budget, there are zonal asymmetries from both the OHF (blue
line in Fig.~\ref{fig-era_summary}e), where there is a larger cooling
tendency in the east than the west, and the net surface shortwave radiation,
where there is a less positive tendency in the west Pacific than in
the central and eastern Pacific (red line in
Fig.~\ref{fig-era_summary}e). The evaporative cooling has a local
maximum near the dateline and comparable values in the west and in the
east despite the SST contrast (black line in
Fig.~\ref{fig-era_summary}e).
The near-equator averaging regions of \citet{wills22}
are a useful way to characterize the bulk zonal contrast:
longitude $110^\circ$ to $180^\circ$ E for the west Pacific minus
$180^\circ$ to $100^\circ$ W for the east Pacific, shown in the gray boxes in
Fig.~\ref{fig-era_summary}a,b.
For the surface energy budget terms, there is a $+32 \, \mathrm{W \, m^{-2}}$
contrast in ocean heat flux, a $-21 \, \mathrm{W \, m^{-2}}$ contrast in
net shortwave radiation, a $-8 \, \mathrm{W \, m^{-2}}$ contrast in evaporative
cooling, a $-5 \, \mathrm{W \, m^{-2}}$ contrast in sensible heat fluxes,
and a $+1 \, \mathrm{W \, m^{-2}}$ contrast in net longwave radiation.
In the approximate surface energy balance calculations that follow, a
characteristic magnitude west-minus-east contrast that is sufficient to bring SST changes
in line with observations is $\approx 3 \, \mathrm{W \, m^{-2}}$, an $\approx 10\%$ perturbation
relative to the climatological contrast in ocean heat flux for $0.5  \, \mathrm{K}$
of basin-mean warming.


One can also evaluate the linear trends in the surface energy
components in ERA5 over the same period. However, there are systematic
changes in time in the surface humidity analysis increments in ERA5
\citep[Fig.~16d of][]{hersbach20}, which likely affect its evaporation
trends. ERA5 has large increasing trends in evaporative cooling (not
shown) that would need to be balanced by correspondingly large
positive energy trend from the ocean heat flux, according to the
residual calculation approach. This is, however, at odds with Pacific
ocean state estimates which have substantial regions of near
thermocline cooling \citep{jiang24}. Given this concern, these
trends are not presented here.


\subsection{Approximate Surface Energy Budget Calculations}

Here, I pursue perturbation surface energy balance calculations to
systematically experiment with OHF, radiative forcing, radiative
feedback, relative humidity changes, and surface wind speed. 
These calculations vary only in longitude $\lambda$ and
the climatological inputs use the same near-equator ($\pm 5^\circ$)
average in latitude as Figure~\ref{fig-era_summary}.  There are five
terms on the right-hand side of the surface energy balance
\eqref{eqn-sfc_en} to represent in terms of a single prognostic sea
surface temperature $SST$ variable.  This section describes the
assumptions used to approximate the terms and specify the relevant
inputs used to solve for the equilibrium surface energy budget and
determine the resulting SST.

The ocean heat flux, OHF, is prescribed from that implied by the ERA5
reanalysis climatology (Fig.~\ref{fig-era_summary}e).  This can be
perturbed by enhancing or suppressing its zonal asymmetry: $\Delta \mathcal{O} = \gamma (\mathcal{O}
- \widehat{\mathcal{O}}) + \widehat{\mathcal{O}}$, where $\widehat{\cdot}$ indicates a Pacific-basin zonal mean
and $\gamma$ is a multiplicative factor and $\gamma = 1$ is an unchanged degree of zonal asymmetry. Figure~\ref{fig-forcing_feedback}d shows $\Delta \mathcal{O}$ for $\gamma = 1.1$, a $10\%$ increase in the climatological zonal asymmetry, in blue. The enhanced zonal asymmetry
can be thought of as a strengthening of the climatological ocean circulation,
but it can also be a proxy for differential heat storage or changes in
ocean stratification that influence the mixed layer heat budget.
That is, the calculations are integrated to equilibrium states ($\partial_t SST = 0$),
but the perturbation to the OHF---the net surface
flux---represents the combined effect of ocean heat flux convergence and
local ocean heat storage. 

The radiative forcing $\mathcal{F}$ in these calculations represents the
net surface radiation that is not linearly related to the local surface
warming (i.e., the feedback component of the radiation, which is discussed next). 
It therefore represents well-mixed greenhouse gases, like carbon dioxide, and
other forcing agents like anthropogenic aerosols. 
Carbon dioxide's instantaneous surface radiative forcing is somewhat smaller than
its more widely examined top of atmosphere forcing in the global mean \citep[e.g.,][]{chen23}. 
The surface radiative forcing of
carbon dioxide depends on the water vapor (with surface forcing
decreasing with increasing water vapor owing to the higher
climatological near-surface opacity) and therefore the relatively dry
east Pacific (away from the intertropical convergence zone, ITCZ) has a
larger radiative forcing than the west Pacific \citep{chen23}. In addition,
non-greenhouse gas forcing agents, such as anthropogenic aerosols, can have distinct zonal spatial
patterns. For example, the aerosol-cloud interactions have a locally negative
forcing in the east Pacific, according to the GCMs examined in \citet{soden17}.
Taken together, the simple starting point examined here is a surface radiative forcing $\mathcal{F}$
used here is spatially uniform value of $4 \, \mathrm{W \, m^{-2}}$. This value is larger
than historical surface forcing, but there is ambiguity in what temperature-dependent
processes are included in the feedback parameter (discussed next), and it has the appeal of
provoking $\approx 0.5 \, \mathrm{K}$ of Pacific-mean warming in the approximate surface energy balance calculations for the reference value of surface feedback parameter chosen.
In addition, a perturbation calculation includes a zonal asymmetry in the forcing
that is an idealized function of longitude $\lambda$ 
that consists of two Gaussian lobes of opposite sign: $A \exp[ (\lambda -
  \lambda_w)^2/\sigma^2] - A \exp[ (\lambda - \lambda_e)^2/\sigma^2 ]$,
  with $\lambda_w = 145^\circ$ E, $\lambda_e = 50^\circ$ W, and $\sigma
  = 40^\circ$, with amplitude $A$.
  This functional form is used for multiple idealized perturbation experiments
  and $A$ carries the appropriate unit. For radiative forcing, its unit is $\mathrm{W \, m^{-2}}$
  and its structure can be seen in the cyan line in Fig.~\ref{fig-forcing_feedback}f.


The net radiative surface radiation $R_{SFC}$ is governed by the
climatological radiation $R_0$ and the temperature-dependent departures
given by the surface
feedback parameter $\lambda_{SFC}$: $R_{SFC} = R_0 + \lambda_{SFC}
SST$.  The subscript $SFC$ is retained throughout to emphasize that $\lambda_{SFC}$ is
the temperature-dependent linearization of surface radiation, including how
the overlying atmospheric thermal structure and clouds affect both shortwave and
longwave surface radiation, and to avoid confusion
with  
longitude $\lambda$. This form is analogous to the typical
top-of-atmosphere form representing outgoing longwave radiation $A + B \Delta
T_s$ used in diffusive energy balance models
\citep[e.g.,][]{north81a,merlis18}.  The $R_0(\lambda)$ is determined
such that for the climatological (indicated by $\overline{\cdot}$)
$\overline{SST}(\lambda)$, it is equal to the climatological ERA5 net
surface radiation: $R_0 = \overline{R}_{SFC}(\lambda) -
\lambda_{SFC} \overline{SST}(\lambda)$, where $\overline{R}_{SFC}$ and $\overline{SST}$
are from ERA5.  The default surface radiative
feedback used here is spatially uniform value of $\lambda_{SFC} = -1.8 \,
\mathrm{W \, m^{-2} \, K^{-1}}$, inspired by typical top-of-atmosphere
values.
Different combinations of $\mathcal{F}$ and $\lambda_{SFC}$ can yield
the observation-like value of Pacific-mean warming of
$0.5 \, \mathrm{K}$, and there is ambiguity about
what physical processes are included in each. In particular, the downward
longwave flux to the surface changes with the temperature-dependent
atmospheric water vapor concentration and cloud properties, but
these are not necessarily proportional to the local SST change \citep[see, for example, the
  analysis of][]{shakespeare22}. Including those on the forcing side of the decomposition,
naturally, means the forcing should be higher than a typical instantaneous surface forcing
value. 
Furthermore, a low feedback and high
forcing (or vice versa) can capture the same net effect for a given climate perturbation
and the radiative restoring of the surface energy balance is generally smaller than 
evaporative restoring in the tropics.
For a zonally asymmetric feedback parameter calculation, 
the same positive and negative Gaussian lobes introduced previously are used.
For the spatially varying $\lambda_{SFC}$, $R_0$ is
recalculated to match the ERA5 climatology.

The latent cooling of evaporation is governed by the bulk aerodynamic
formula: $LE = L C_k \rho |\mathbf{u}| [ q^*(SST) - q_a] \approx L C_k
\rho |\mathbf{u}| [ q^*(SST) - \mathcal{RH}q^*(T_a) ]$, with the
latent heat of vaporization $L$, turbulent drag coefficient for enthalpy $C_k$, air density $\rho$,
surface wind speed $|\mathbf{u}|$, saturation
specific humidity $q^*$, surface-air specific humidity $q_a$, 
surface-air relative humidity $\mathcal{RH}$, and surface-air temperature
$T_a$. The parameter values are $C_k = 1.1 \times
10^{-3}$, $\rho = 1.3 \, \mathrm{kg \, m^{-3}}$, and $L = 2.5 \times
10^{6} \, \mathrm{J \, kg}$.
The control $|\mathbf{u}|$ is the
climatological average of the ERA5 $10$-m wind speed computed from
$6$-hourly output.  The climatological ERA5 $\mathcal{RH}$, including its spatial
structure, is calculated from monthly mean $2$-m temperature and dewpoint
using the ECMWF formulation for the saturation vapor pressure 
\citep{simmons99}. To determine $T_a$, the ERA5
climatological air--sea temperature difference is added to the prognostic
$SST$.  Perturbation calculations where $\mathcal{RH}$  and $|\mathbf{u}|$
are varied (Fig.~\ref{fig-rh_wind}a,b) can be thought of as
temperature-dependent changes in the surface wind and relative
humidity.  The Clausius-Clapeyron rate sensitivity of $q^*(T)$ is
the dominant temperature-dependent term in the calculations
presented here. It is also the only non-linearity, so one could
linearize it to solve for the perturbation energy balance without numerical integration
\citep[as in][]{zhang14b}, but the non-linear form is retained here.


The air--sea temperature difference $SST-T_a$ modulates the evaporative flux.
Here, the climatology is prescribed from ERA5 and left unchanged in
perturbation calculations. The air--sea temperature disequilibrium does change
in simulations of global warming \citep[e.g.,][]{richter08}.
However, it is not one of the isolated
perturbation cases performed here. 
This is because 
changing air--sea disequilibrium has less leverage than the surface $\mathcal{RH}$,
given the sensitivity of the bulk formula and is a small term in the
GCM energy budget decompositions presented in \citet{zhang14b} and \citet{hwang17}. It may warrant further consideration
for historical changes.

The turbulent sensible flux is lumped with the surface radiation in the calculations that follow.
The 
$R_0$ formula that includes the climatological $SH$ is $R_0(\lambda) = \overline{R}_{SFC}(\lambda) + \overline{SH} - \lambda_{SFC} \overline{SST}(\lambda)$, with all
overline quantities from ERA5. This implies the surface feedback parameter
 $\lambda_{SFC}$ 
is capturing the combined SST dependence of $R_{SFC}$ and $SH$. This choice is a detail, however.
Despite the importance of the change
in sensible flux for the tropical-mean or
global-mean surface energy budget and concomitant role in the hydrological
sensitivity \citep{boer93,schneider10}, it is not large in
absolute terms nor in east--west contrast
(Fig.~\ref{fig-era_summary}e). Furthermore, GCM-simulated changes in 
the east--west contrast of sensible fluxes in the tropical Pacific
are small \citep{dinezio09,zhang14b}.
Last, it is known on physical grounds that the SST dependence of sensible heat fluxes are weak compared to the latent energy flux
because of the smallness of the Bowen ration for tropical temperatures
\citep[e.g.,][]{hartmann94b,merlis10}.
With this approach to surface radiation and sensible heat fluxes defined,
the linearized surface energy budget analysis of \citet{hwang17} is a useful point of
comparison. There, they 
considered $SH$ changes in isolation and separated the radiation into shortwave cloud
and clear vs. longwave, rather than a combined shortwave and longwave forcing
vs. feedback. They also presented changes in air--sea disequilibrium,
for a $7$-term decomposition (vs. $5$ inputs varied here). 

The ERA5 inputs are smoothed from the native
$25$-km resolution over a running $20$ grid-point window in longitude.
I numerically calculate the equilibrium energy budget by
time-integrating to equilibrium using a heat capacity of water of
$1 \, \mathrm{m}$ depth for $C$ over $600$ days. A more Earth-like
mixed-layer depth integrated to equilibrium with a longer period
of integration yields the same results.

To summarize, the approximate surface energy balance is
\begin{equation}
\begin{split}
  C \partial_t SST = &R_0 + \lambda_{SFC} SST + \\ 
                     &L C_k \rho |\mathbf{u}| \left[ q^*(SST) - \mathcal{RH}q^*(T_a) \right] + \mathcal{O} + \mathcal{F},
  \label{eqn-sfc_en_approx}
\end{split}
\end{equation}
with symbols defined above.
The combination of $R_0 + \lambda_{SFC} SST$ is labeled $\lambda \Delta SST$ in the
figure legends, though it includes $SH$ as previously stated.
There are five individual perturbations to this approximate energy balance that are examined next (section~\ref{sec-ind_pert}) to determine the magnitude required to roughly capture the east--west warming magnitude:
\begin{enumerate}
  \item The climatological zonal asymmetry in $\mathcal{O}$ is enhanced.
  \item Idealized zonal asymmetry is added to the surface radiative forcing $\mathcal{F}$.
  \item Idealized zonal asymmetry is added to the surface radiative feedback $\lambda_{SFC}$.
  \item Idealized zonal asymmetry is added to the climatological $\mathcal{RH}$.
  \item The climatological zonal asymmetry in $|\mathbf{u}|$ is enhanced.
\end{enumerate}  
In these individual perturbation calculations, the magnitude of the
perturbation are adjusted to reach the target west-minus-east contrast
of $0.4 \, \mathrm{K}$, inspired by observed trends
(Fig.~\ref{fig-era_summary}b,d). In all individual perturbation
calculations, all of the zonal asymmetries of the climatology are
included and a single, isolated additional perturbation zonal
asymmetry is added.  A combined perturbation calculation with all five
perturbed at weaker amplitude and a simultaneous increase in the zonal
asymmetry of $\mathcal{O}$ and $|\mathbf{u}|$ are also presented
(Fig.~\ref{fig-combined_perturbations} in section~\ref{sec-comb_pert}).

\section{Individual Perturbations} \label{sec-ind_pert}

\subsection{Climatology of Approximate Surface Energy Balance}

Figure~\ref{fig-era_summary}c shows the equilibrated surface
temperature for the approximate surface energy balance \eqref{eqn-sfc_en_approx} calculation
with no radiative forcing (black dotted line).  The basic structure of the tropical Pacific
temperature is captured, with warmer SST in the west than the east,
and there is an $\approx 1 \, \mathrm{K}$ warm bias in the west and a
somewhat smaller cold bias in the central Pacific. The
lack of time-dependent correlations between the imposed fields (i.e.,
fluctuations in wind speed, surface-air temperature, and surface-air
relative humidity are not independent) is a likely reason for
biases. Using the averaging regions of \citet{wills22}, there is a
$3.85 \, \mathrm{K}$ west-minus-east SST difference vs. ERA5's $2.75 \, \mathrm{K}$.
The surface energy
balance reproduces the ERA5 climatology by construction (not
shown).

\subsection{Control Response: Evaporative Damping}

\begin{figure*}[h!]
  \noindent\includegraphics[width=32pc,angle=0]{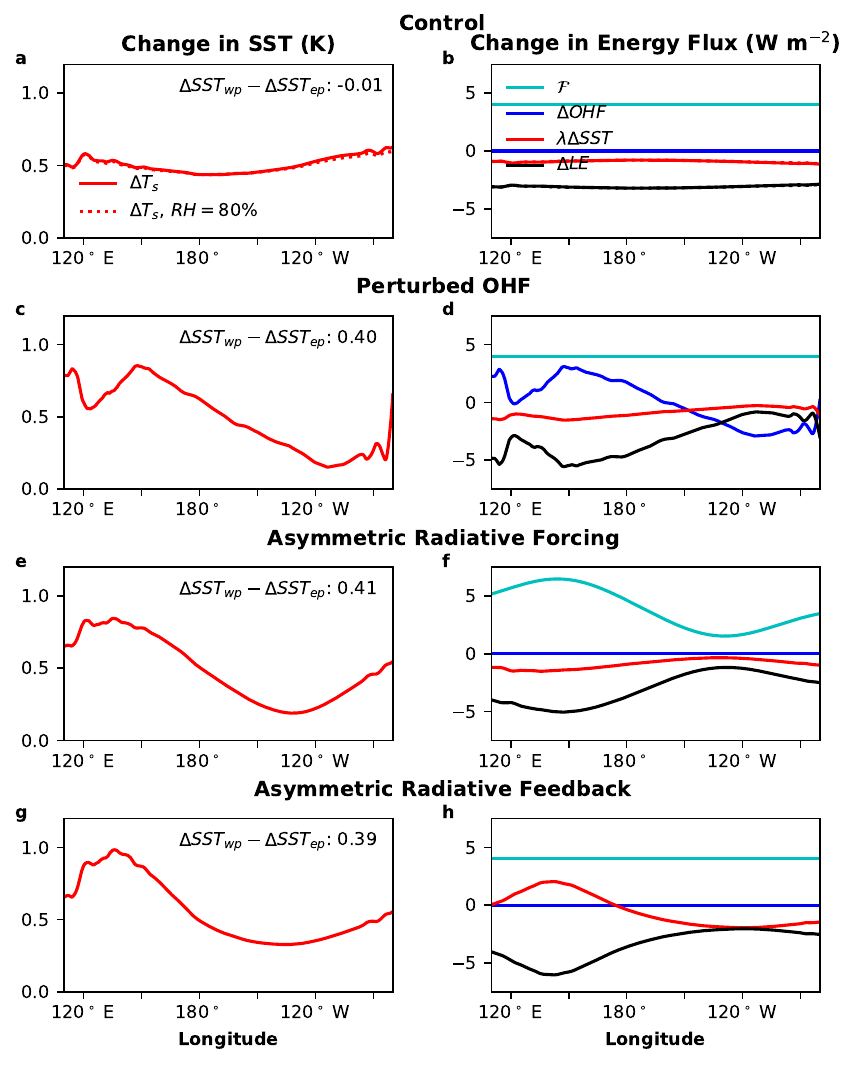}\\
  \caption{(left) Change in SST (units $\mathrm{K}$) and (right)
    change in surface energy tendencies (units $\mathrm{W \, m^{-2}}$)
    for approximate surface energy balance calculations.  (a,b)
    Calculations assuming uniform forcing, feedback, unchanged ocean
    heat flux, unchanged relative humidity, and unchanged surface
    wind speed with dotted lines showing the results for a spatially
    uniform relative humidity of $80\%$. As above, except for (c,d)
    strengthened ocean heat flux, (e,f) zonally asymmetric radiative
    forcing, and (g,h) zonally asymmetric radiative
    feedback.}\label{fig-forcing_feedback}
\end{figure*}

Figure~\ref{fig-forcing_feedback}a shows the equilibrated surface
temperature for a spatially uniform forcing with unchanged ocean heat
flux and spatially uniform feedbacks. There is enhanced warming in the
east, a local minimum in the central Pacific, and a west-minus-east
contrast that decreases with warming. The magnitude of the decrease is
$0.013 \, \mathrm{K}$ for a Pacific-mean warming of $0.5  \, \mathrm{K}$,
a $\approx 1\% \, \mathrm{K^{-1}}$ reduction relative to the
climatology of a contrast of $3.7 \, \mathrm{K}$.
This is weaker than the evaporative damping scaling, which has a surface Clausius-Clapeyron
rate of $\approx 6\% \, \mathrm{K}^{-1}$

The individual components of surface energy balance are nearly
spatially uniform (Fig.~\ref{fig-forcing_feedback}b).  There is no
change in the ocean heat flux (blue), the forcing is spatially uniform
(cyan), the surface radiation change has a slight inhomogeneity from
the pattern of SST warming (red), and the latent cooling has an offsetting
slight inhomogeneity (black).
The basic balances of a spatially uniform increase in evaporative
cooling implying a spatially inhomogeneous SST change, with enhanced
warming in the cold region, is consistent with evaporative damping.
The mildly enhanced west Pacific warming is, at first glance,
inconsistent with the Clausius-Clapeyon sensitivity of the evaporative
damping mechanism. However,
the climatological wind speed is lower (Fig.~\ref{fig-rh_wind}b) and, all else equal, this
implies more warming is needed to balance the radiative forcing.
This can be readily seen if one assumes a two-term perturbation
energy balance between the radiative forcing and linearized evaporative
cooling:
\begin{equation}
  \Delta SST \approx \frac{\mathcal{F}}{L \rho c_k |\mathbf{u}| (1 - \mathcal{RH}) \partial_T q^*},
\label{eqn-two_term_bal}
\end{equation}
where the air--sea temperature difference is neglected for simplicity.


Before turning to zonal asymmetries in forcing and other inputs, it is
worth examining the role that the imposed climatological relative
humidity plays in determining the response (dotted lines in Fig.~\ref{fig-forcing_feedback}a,b).
The surface-air relative humidity
varies between $80\%$ and $84\%$ with a local minimum near the
dateline and a maximum in the east Pacific
(Fig.~\ref{fig-rh_wind}a).
\citet{seager19} found little difference between $80\%$ $\mathcal{RH}$ and ERA-Interim
$\mathcal{RH}$ in their intermediate complexity coupled atmosphere--ocean model simulations,
and this is what is seen here, though they found sensitivity to using the climatological CMIP5 $\mathcal{RH}$.
The dotted lines in Figure~\ref{fig-forcing_feedback}a show a modest reduction in warming
in the west and east where the $\mathcal{RH}$ was reduced---the sense of the change
anticipated by \eqref{eqn-two_term_bal}. The surface energy budget changes are nearly indistinguishable (Fig.~\ref{fig-forcing_feedback}b).




\subsection{Strengthened Ocean Heat Flux}

One can provoke enhanced warming in the west over the east by
increasing the zonal asymmetry of the climatological ocean heat flux
(blue line Fig.~\ref{fig-era_summary}e), which is a relative warming
in the west and a relative cooling in the east compared to the
Pacific-mean cooling tendency (blue line in
Fig.~\ref{fig-forcing_feedback}d). This perturbation can be thought of
as either an ocean thermostat-type transient change or an equilibrium
strengthening of the divergence/convergence of ocean heat transport
(e.g., from stronger trade winds).  Here, the zonal asymmetry in the
ERA5 OHF is rescaled by $+10\%$ as a perturbation to the surface
energy balance calculation, which has a $\approx 30 \, \mathrm{W \,
  m^{-2}}$ climatological west--east contrast.  The resulting
structure in SST change has a clear west--east contrast
(Fig.~\ref{fig-forcing_feedback}c), with some detailed structure
arising from the OHF climatology that is being rescaled.  Given that
the radiative forcing provokes $0.5 \, \mathrm{K}$ of Pacific mean
warming, this implies a $\approx 20\% \, \mathrm{K^{-1}}$ sensitivity
in the ocean heat flux is the magnitude change if the OHF is a
temperature-dependent response to forcing. The dominant way the energy
balance re-equilibrates to this imposed zonal asymmetry is via
evaporative cooling (black line in Fig.~\ref{fig-forcing_feedback}d).
Here and in all of the perturbation calculations that follow, there is
little discussion of the more detailed regional structure (e.g., the
structure here near $120^\circ$ E or near the eastern edge of the
basin). This is because of the primary interest in developing the
order-of-magnitude knowledge of the sensitivities for the bulk
west--east contrast and because some of this structure results from
the particulars of the imposed perturbations, which are highly
idealized. To that end, Figure~\ref{fig-summary}b summarizes the
change in the west-minus-east energy budget terms, highlighting the
approximate two-term balance where the increase in zonal asymmetry
of the OHF is met by a corresponding opposing zonal asymmetry in
evaporative cooling.

\subsection{Spatially Varying Radiative Forcing}

One can provoke enhanced warming in the west over the east with a
spatially varying radiative forcing that is larger in the west than
the east. This perturbation can be thought of as probing the
magnitude of uncertainty in the spatial pattern of anthropogenic
radiative forcing, such as aerosol-cloud interactions, that
would be quantitatively important to the east--west contrast
in SST trends.
Figure~\ref{fig-forcing_feedback}e,f shows the results of
surface energy balance calculations with the same Pacific-mean
forcing, but with this zonal asymmetry (cyan line).  There is $3.3
\, \mathrm{W \, m^{-2}}$ more forcing in the west than the east (Fig.~\ref{fig-summary}b) and
this provokes $0.4 \, \mathrm{K}$ more warming. The imposed zonal
asymmetry in surface radiative forcing is largely balanced by
asymmetry in the evaporative cooling, with a secondary role for the
radiative cooling (black and red lines respectively in
Fig.~\ref{fig-forcing_feedback}f).
The magnitude of the OHF zonal contrast in the previous perturbation
calculation was $3.2 \, \mathrm{W \, m^{-2}}$,
and it is unsurprising that this closely matches the contrast for the radiative
forcing: they are equivalent imposed tendencies in these calculations
\eqref{eqn-sfc_en_approx}, albeit
with different spatial structure.

\subsection{Spatially Varying Radiative Feedback}

One can provoke enhanced warming in the west over the east with a
spatially varying radiative feedback that is more amplifying (i.e., more destabilizing or
less negative with the typical feedback sign convention)
in the west than the east. In other words, the local surface feedback
determines the strength of the radiative restoring of a surface
temperature anomaly and one can suppress warming in the east
with a more stabilizing feedback parameter or enhance warming in the
west with a less stabilizing feedback parameter.
This perturbation can be thought of as probing what spatial structure
and amplitude of uncertain cloud radiative feedbacks (such as a stabilizing
low-cloud feedback in the east Pacific or an amplifying high-cloud
feedback in the west Pacific) would affect the east--west SST trends.
Given that net
surface radiation is a subdominant means of re-balancing the surface
energy budget, a large zonal asymmetry in the surface feedback
parameter is required to reach the target temperature
contrast.  Figure~\ref{fig-forcing_feedback}g,h shows the results of
surface energy balance calculations with the same Pacific-mean
feedback, but zonal asymmetry of the same spatial structure as the
radiative forcing in the previous perturbation calculation.  There is a $5.3 \, \mathrm{W \, m^{-2} \, K^{-1}}$ peak-to-trough
contrast in the surface radiative feedback parameter ($\approx 4 \, \mathrm{W \, m^{-2} \, K^{-1}}$ using the averaging regions) to provoke
$\approx 0.4 \, \mathrm{K}$ more warming in the west.  Relative to the
zonal mean $\lambda_{SFC} = - 1.8 \, \mathrm{W \, m^{-2} \, K^{-1}}$,
this is a positive, locally unstable surface radiative feedback
in the west: there is less surface radiative
cooling with warmer SST (red line above zero in the west in
Fig.~\ref{fig-forcing_feedback}h).
Local feedback can be unstable (i.e., positive) in certain regions in
comprehensive GCMs \citep[particularly where surface albedo changes,
  e.g.,][]{feldl23}.
The zonal asymmetry from surface
radiative cooling is balanced by the latent cooling (black line in
Fig.~\ref{fig-forcing_feedback}h). The equilibrated zonal asymmetry in
evaporative cooling is about a $3 \, \mathrm{W \, m^{-2}}$ contrast,
as in the case with zonal asymmetry in surface radiative forcing (Fig.~\ref{fig-summary}b).

\subsection{Spatially Varying Surface RH Changes}

\begin{figure*}[t]
  \noindent\includegraphics[width=32pc,angle=0]{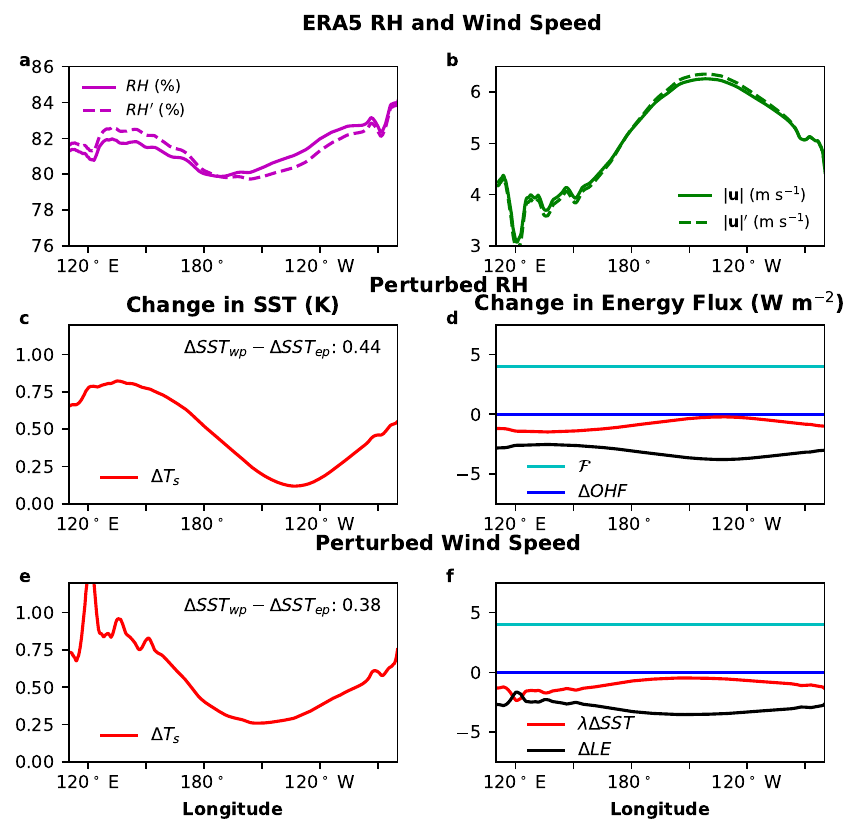}\\
  \caption{Equatorial ocean-only Pacific climatology of (a) surface RH (solid
    magenta, units $\%$) with idealized perturbed surface RH (dashed magenta)
    and (b) surface wind speed (solid green, units $\mathrm{m \, s^{-1}}$) with its
    perturbed surface wind speeds of an $8\%$ increase in its climatological
    zonal asymmetry (dashed green)
    in ERA5 averaged within $5^\circ$ of the equator.  Change in SST and change in surface
    energy tendencies for approximate surface energy balance
    calculations as in the control (Fig.~2a,b), except for (c,d) zonally asymmetric
    relative humidity change and (e,f) strengthened zonally
    asymmetric wind speed. }\label{fig-rh_wind}
\end{figure*}

One can provoke enhanced warming in the west over the east by
preferentially increasing the relative humidity in the west and
decreasing it in the east.
Though GCM simulations have fairly spatially uniform surface-air
relative humidity changes over the ocean \citep[e.g.,][]{ogorman10},
a modest change can exert a substantial influence of the perturbation
surface energy budget. Absent strong theoretical constraints
on surface relative humidity changes, it is difficult to rule out
errors in GCM-simulated spatial trends. 
As an idealized approach, one can use the same simple
zonally asymmetric function that was 
used for the radiative forcing to perturb the $\mathcal{RH}$ (dashed magenta line in Fig.~\ref{fig-rh_wind}a).

Figure~\ref{fig-rh_wind}c shows the change in SST for this idealized
zonally asymmetric change in relative humidity.
To have a $0.4 \, \mathrm{K}$ increase in SST contrast, a west--east
relative humidity contrast of
about $0.5 \%$ is required, $\approx 1 \%  \, \mathrm{K^{-1}}$ sensitivity
for the $\mathcal{RH}$ contrast.
By increasing the
relative humidity in the west, a larger warming is needed to re-balance
the imposed forcing by evaporative cooling. In having a larger warming, 
associated with the muted increase in evaporation,
the radiative restoring can become an important means of re-equilibrating
the local surface energy balance (black vs. red line in the west of Fig.~\ref{fig-rh_wind}d,
Fig.~\ref{fig-summary}b).
Analogously, decreasing the relative humidity in the east Pacific
implies that a modest warming can give rise to the necessary evaporative
cooling to balance the surface radiative forcing. 
Interestingly, the equilibrated spatial contrast in the change in evaporative
cooling is muted compared to the cases shown in Figs.~\ref{fig-forcing_feedback}d,f,h.


\subsection{Strengthened Surface Wind Speed Zonal Asymmetry}

One can provoke enhanced warming in the west over the east by
preferentially decreasing the surface wind speed in the west and
increasing it in the east.
This storyline can be thought of as a particular form of
a large-scale circulation change that affects the surface wind speed
(a weakening of divergent circulations in the west and a strengthening in
the east),
but it could also reflect changes in unresolved processes,
such as convective gustiness, that may be uncertain in GCM
simulations of climate change.
A wind speed change would also provoke a change in ocean circulation,
and this particular combined perturbation is assessed in the next section.
An overall increase in wind speed tends to reduce the SST, so an
enhancement of the climatological zonal asymmetry in the wind speed
(dashed green line in Fig.~\ref{fig-rh_wind}b)
is applied to keep the Pacific-mean temperature change similar.
Figure~\ref{fig-rh_wind}e shows the change in SST for an idealized
zonally asymmetric change in surface wind speed, an $8\%$ increase
in the climatological zonal asymmetry---decreasing in the west and increasing
the wind speed to the east of the dateline.
This pathway toward enhanced warming in the west is reminiscent of the
way in which the warm percentiles of the Pacific SST undergo warming during the
interannual variability of the El Ni\~no--Southern Oscillation (ENSO), when 
there is warming associated with a reduction in surface wind speed during El-Ni\~no events.
\citep{hogikyan22}.
The effect of a decrease of wind speed in the west and an increase
in the east is to decrease the evaporative cooling in the west
and increase it in the east relative to the case of unperturbed
wind speed (black line in Fig.~\ref{fig-rh_wind}f vs. Fig.~\ref{fig-forcing_feedback}b).
With relatively less evaporative cooling in the west, there is a need
for more warming so that the radiative restoring can re-equilibrate the
surface energy budget (red line in Fig.~\ref{fig-rh_wind}f, Fig.~\ref{fig-summary}b).

\section{Combined Perturbations} \label{sec-comb_pert}

To this point, individual factors affecting surface energy balance calculations
were assessed in a one-at-a-time manner.
A mathematical way of conceptualizing what these calculations tell us
is to consider the west--east contrast in SST $\delta_{w-e} SST$
in terms of the
input variables presented: $\delta_{w-e} SST (\mathcal{O}^{asym},
\mathcal{F}^{asym}, \lambda_{SFC}^{asym}, \mathcal{RH}^{asym},
|\mathbf{u}|^{asym})$, where the superscript $asym$ makes clear that
particular spatially asymmetric forms of these variables were
assessed. One can perform a Taylor-series
expansion of $\delta_{w-e} SST$ for a given global-mean or Pacific-mean surface
temperature change $\langle T \rangle$:
\begin{equation}
  \frac{d \delta_{w-e} SST}{d \langle T \rangle} \approx \frac{ \partial \delta_{w-e} SST}{ \partial \mathcal{O}^{asym}}\frac{ \partial \mathcal{O}^{asym}}{\partial \langle T \rangle} +  \frac{ \partial \delta_{w-e} SST}{ \partial \mathcal{F}^{asym}}\frac{ \partial \mathcal{F}^{asym}}{\partial \langle T \rangle} + \ldots
  \label{eqn-Taylor_expansion}
\end{equation}  
The first-order sensitivities were identified in these calculations and there are, of
course, higher order terms (e.g., quadratic) in individual
perturbations and cross-terms that were neglected.

Higher-order
terms in Taylor-series expansion are smaller in magnitude by
definition, and these perturbations are $\approx 10\%$ of the climatological
energy budget because the perturbations have contrasts of $\approx 3 \, \mathrm{W\, m^{-2}}$
and the climatology has contrasts of $\approx 30 \, \mathrm{W\, m^{-2}}$.
Bolstering the claim that a first-order linearized approach is adequate is 
the diagnostic surface energy budget analysis in GCM simulations
of \citet{hwang17}.
However, it is worth examining simultaneous perturbation cases for the calculations
presented here. First, a combined case with each of the individual five factors previous assessed
is presented. Then, a simultaneous ocean heat flux and wind speed case is presented.

\subsection{Simultaneous Perturbations}

\begin{figure*}[t]
  \noindent\includegraphics[width=32pc,angle=0]{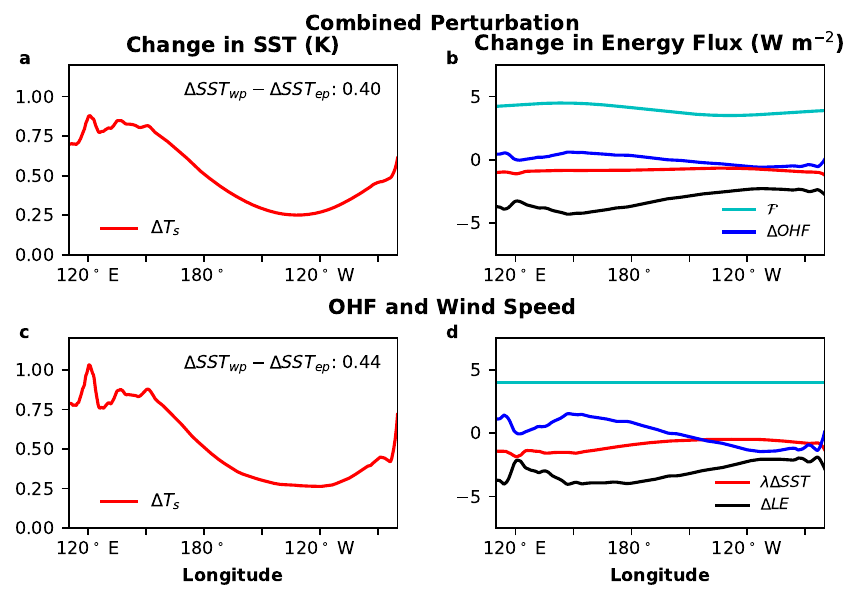}\\
  \caption{(left) Change in SST (units $\mathrm{K}$) and (right)
    change in surface energy tendencies (units $\mathrm{W \, m^{-2}}$)
    for approximate surface energy balance calculations.  (a,b)
    Calculations for a combined perturbation with each
    factor perturbed by $20\%$ of the magnitude of the individual
    perturbation calculations shown in Figs.~2,3.
    (c,d) Calculations for a combined perturbation with both ocean
    heat flux and wind speed perturbed, where the climatological
    zonal asymmetry in each is increased by $5\%$.}\label{fig-combined_perturbations}
\end{figure*}

\begin{figure*}[t]
  \noindent\includegraphics[width=24pc,angle=0]{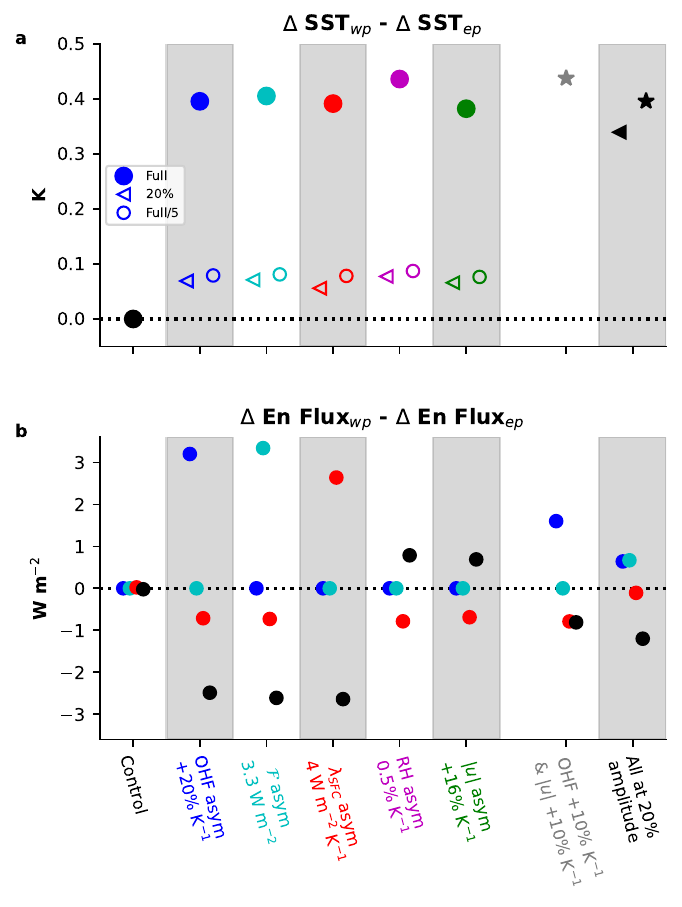}\\
  \caption{A summary of the perturbation calculations for (a) the
    west--east SST contrast (in $\mathrm{K}$) and (b) the 
    change in west-minus-east surface energy tendencies (in $\mathrm{W \, m^{-2}}$).
    Linearity is examined by comparing individual perturbation
    calculations with $20\%$ amplitude (triangles in panel a, with their sum in black),
    to the full response scaled down by the same factor (open circles). 
}\label{fig-summary}
\end{figure*}

Having established individual sensitivities, it is important to
determine the extent to which these surface energy balance
calculations are linearly additive. Here, the five perturbations that
met the simple SST west--east contrast target are rescaled by a
factor of five ($20\%$ of each) and simultaneously perturbed.
Figure~\ref{fig-combined_perturbations}a shows the change in SST for the combined
perturbation.  It has a $0.4 \, \mathrm{K}$ SST contrast, indicating a
linear response by this metric.  Figure~\ref{fig-combined_perturbations}b shows the
change in surface energy tendencies. The predominant asymmetries are
the relatively positive tendencies in the west in
radiative forcing and OHF, as prescribed in the calculation. These are
then balanced by a relatively enhanced evaporative cooling in the west
over the east (black line in Fig.~\ref{fig-combined_perturbations}b, Fig.~\ref{fig-summary}b).

To further assess linearity, Figure~\ref{fig-summary}a shows the results of
individual perturbations at $20\%$ amplitude (triangles) and the full amplitude
individual perturbations scaled down by $20\%$ (open circles). There is a modest amount
of amplitude dependence to the response, with the weak amplitude forcing having a smaller
change in the SST contrast metric than the rescaled full response.
There is also a $\approx 15\%$ smaller response when comparing the sum of small amplitude perturbations (black triangle) and the simultaneous, combined five perturbations at that amplitude (black star). 
In summary, the combined perturbation approximately follows a superposition of
the individual perturbations and these calculations are close to linear over the range of
forcing amplitudes examined here.

Here, it is important to emphasize that
this demonstration of linear superposition does not imply that
each of these perturbations at this particular amplitude is a
``realistic'' scenario to account for the historical response of the
Pacific SST. It is natural to expect some of these perturbations to
have correlated amplitudes
because they are dynamically related.

\subsection{Ocean Heat Flux and Wind Speed}

One particular combination that is of interest is the change in zonal
asymmetry of the ocean heat flux and surface wind speed.  The widely
discussed Bjerknes feedback of the ENSO cycle has a reinforcing
interaction between the SST gradient, wind stress, and ocean heat
flux.  A climatological version of the Bjerknes feedback may affect
the sensitivity of the tropical Pacific climate to radiatively forced
changes. This combined perturbation is also potentially relevant to
decadal timescale internal variability generated by atmosphere--ocean coupling.

To quantify a storyline for a climatological Bjerknes feedback, the
perturbation wind speed can be proportional to the perturbation
OHF. I note, however, that it is not inevitable that these two
strongly constrain each other as the change in wind speed with
climate includes factors beyond the mean state east-west overturning
circulation that zonal asymmetry in the OHF strongly influences.
The wind speed entering the bulk formula can vary with other aspects
of the atmospheric flow, including its transient components.

Figure~\ref{fig-combined_perturbations}c shows the change in SST for the combined
OHF and wind speed zonal asymmetry perturbation, where each has its climatological
asymmetry increased by $5\%$.  It has a $0.44 \, \mathrm{K}$ SST contrast.
This perturbation is half that as the isolated OHF case and slightly more than half the
isolated wind speed case (Fig.~\ref{fig-rh_wind}e shows an $8\%$ increase) and the response in the SST metric is about $10\%$ larger. Again,
it is clear that the surface energy balance is close to linear in perturbations of this magnitude,
providing an a posteriori justification for considering isolated perturbations even though the
climate system has responses in these variables that are tied together physically.
  Figure~\ref{fig-combined_perturbations}d shows the
change in surface energy tendencies for this case. The predominant asymmetries are
between the OHF and evaporative cooling.

To return to the Taylor-series expansion formalism, I can present the
results of this calculation as either perturbing two temperature-dependent
factors simultaneously,
\begin{equation*}
 \frac{d \delta_{w-e} SST}{d \langle T \rangle} \approx \frac{ \partial \delta_{w-e} SST}{ \partial \mathcal{O\
   }^{asym}}\frac{ \partial \mathcal{O}^{asym}}{\partial \langle T \rangle} + \frac{ \partial \delta_{w-e} SST}{ \partial |\mathbf{u}|^{asym}}\frac{ \partial |\mathbf{u}|^{asym}}{\partial \langle T \rangle}
\end{equation*}
or as a perturbation where the wind speed change depends on the ocean heat flux change,
\begin{equation*}
\frac{d \delta_{w-e} SST}{d \langle T \rangle} \approx \frac{ \partial \delta_{w-e} SST}{ \partial \mathcal{O}^{asym}} \frac{ \partial \mathcal{O}^{asym} }{\partial \langle T \rangle} + \frac{ \partial \delta_{w-e} SST}{\partial |\mathbf{u}|^{asym} } \frac{\partial |\mathbf{u}|^{asym} }{\partial \mathcal{O}^{asym} } \frac{\partial \mathcal{O}^{asym} }{\partial \langle T \rangle}.
\end{equation*}
In the latter case, an $\approx 3\%$ change in asymmetric wind speed per $\mathrm{W \,
  m^{-2}}$ of asymmetry in OHF---the middle partial sensitivity of the second term---is equivalent to the case presented
because the $5\%$ increase in OHF translates to $1.6 \, \mathrm{W \,
  m^{-2}}$ of asymmetry in it. This $\approx 3\%$ sensitivity in
asymmetric wind speed per $\mathrm{W \, m^{-2}}$ of asymmetry in OHF has
a
physical basis: it is 
the magnitude that is needed to zero the asymmetry in wind speed if the
OHF became symmetric, given the climatological zonal contrast of $\approx 30 \,
\mathrm{W\, m^{-2}}$.
Alternatively, one can estimate this sensitivity by regressing individual years
of the west-minus-east indices to capture interannual variability (e.g., from ENSO),
and there is an $\approx 1\%$ change in $ |\mathbf{u}|^{asym}$ per
$\mathrm{W\, m^{-2}}$ of $\mathcal{O}^{asym}$ in this case.

In summary, the combined perturbation cases presented in this section confirm
that the perturbation surface energy budget calculations are close to linear for
the magnitude of perturbations under consideration here. They can also be viewed
as representing physically interacting components of the tropical climate, such as
the asymmetry of the wind speed and ocean heat flux---inspired by the ENSO cycle.


\section{Discussion and Conclusions} \label{sec-conclusion}

Here, idealized surface energy budget calculations were performed with
various imposed factors. These factors include the zonal asymmetries
in the ocean heat flux, surface radiative forcing and
feedback, surface-air relative
humidity, and surface wind speed. The magnitude of the different perturbations
sufficient to bring SST changes in line with observed trends were determined.

The perturbation calculations can be thought of in
terms of a series expansion of the surface energy budget \eqref{eqn-Taylor_expansion}.
Therefore, these calculations are
adjacent
to feedback analysis.  While one can perform a formal feedback
analysis for surface energy budgets,
this avenue was not pursued for
two reasons.  First, the system examined here has a spatial dimension
and this increases the space of possible perturbations.  There are
higher-dimension feedback calculations performed for GCM simulations of
Earth's atmosphere and surface
that use the model-simulated responses; however the motivation here
is that climate model simulated responses may be inconsistent with
Earth's historical changes. So, it is more straightforward to simply
identify the ``storylines'' of perturbations of adequate magnitude to
provoke SST trends using numerical calculations with idealized
zonally asymmetric input.


Discussion of feedback analyses naturally raises the question of the
focus here on the surface energy balance rather than that of the top of atmosphere (TOA).
The TOA perspective is highly effective in the climatic regime where
there is a unique relationship between the surface temperature and TOA
radiative fluxes (particularly via convective coupling). This breaks down in
high latitudes and the distinctive roles for changes in the free
troposphere and surface temperature have been a longstanding aspect of
low-cloud research \citep[e.g.,][]{bretherton13} that is relevant to
the subsiding branch of the Walker circulation in the east Pacific.  Also, anthropogenic
aerosol forcing---a potential pathway to zonally asymmetric forcing---can
have disparate TOA and surface forcing.  In practical
terms, the radiative feedbacks in the calculations here are defined in terms of the surface
radiative fluxes, and these are less widely examined than their
TOA counterparts. Unlike the TOA budget, the surface energy budget has no
explicit atmospheric energy transport term, which means it affects
surface temperature by indirectly influencing the downward longwave
radiation, surface-air relative humidity, and surface wind speed.
These approximate surface energy calculations
allow one to directly assess the possible role of surface relative
humidity and wind speed changes, which have been 
discussed in intermediate complexity coupled models
\cite[][]{seager19}.  Finally, a key virtue is that the isolated role of
the ocean heat flux changes on the surface temperature can be readily
evaluated.

What is the magnitude of changes in energy fluxes in the tropical
Pacific that could provoke the recent several decades of observed sea
surface temperature trends?
As a reasonable target for the approximate surface energy balance
calculations, a Pacific-mean warming of $0.5 \, \mathrm{K}$
and a
west--east contrast in change of $0.4 \, \mathrm{K}$ are comparable
with observational estimates.
The mean warming is provoked by the spatially uniform radiative forcing.
To achieve the zonal asymmetry with the individual perturbations
examined here, the following magnitudes
are needed:
\begin{enumerate}
  \item{a $\approx 20\% \, \mathrm{K}^{-1}$ increase in the
  zonally asymmetric component of the ocean heat flux}
  \item{surface
  radiative forcing $\approx 3 \, \mathrm{W \, m^{-2}}$ stronger in
  the west than the east}
  \item{a zonal asymmetry in surface radiative feedbacks of
      $\approx 4 \, \mathrm{W \, m^{-2} \, K^{-1}}$}
  \item{ a zonally asymmetric 
        relative humidity change of $\approx +0.5 \% \, \mathrm{K^{-1}}$ in the west and $\approx -0.5 \% \, \mathrm{K^{-1}}$ in the east}
  \item{or  $\approx 16 \% \, \mathrm{K^{-1}}$  increase in the zonal asymmetry in wind speed.}
\end{enumerate}
A combination of perturbations can also meet the SST contrast
target. In particular, a combination of these five perturbations at
$20\%$ of their individual amplitude does---showing the amplitude is
small enough that linear superposition approximately holds.  Finally, a combined
$\approx 10\% \, \mathrm{K}^{-1}$ change in the zonal asymmetry of the
ocean heat flux and wind speed meets the SST contrast target.

The individual sensitivities identified are large in magnitude.
Note that the per
K in the sensitivities quantified used the Pacific-basin mean (i.e.,
local) SST change, and rescaling to global-mean surface temperature
change would decrease these magnitudes. 
A common benchmark in quantifying temperature-dependent climate changes is that of
the Clausius-Clapeyron relation of $\approx 6\% \, \mathrm{K}$ for tropical temperatures
and normalizing by local SST change rather than the global-mean is appropriate.
This rate is a useful way of delineated high sensitivity changes, as it is both a
robust and rapidly change aspect of the climate.
According to these perturbation surface energy balance calculations,
the changes in zonal asymmetries that are sufficient to generate SST trends comparable
to observations exceed the Clausius-Clapeyron rate.

The equilibrated surface energy budgets for these different perturbations
are also interesting.
Broadly, there are two ways in which the surface energy budget rebalances.
First, in the case of perturbed ocean heat flux, zonally asymmetric radiative
forcing, or zonally asymmetric radiative feedback, the imposed changes
have enhanced positive tendencies in the west compared to the east 
that are balanced by enhanced evaporative cooling in the west compared to the east (Fig.~\ref{fig-summary}b).
Second, in the case of perturbed relative humidity and wind speed,
these perturbations drive enhanced warming in the west because they are
designed to preferentially reduce the evaporative cooling in that
region, leaving  the temperature-dependent radiative restoring to balance the
differential warming in the west (Fig.~\ref{fig-summary}b). The combined case behaves like the
first series of calculations, with more evaporative cooling in the west
than the east. 

How does a deeper understanding of the surface energy balance response
to perturbations aid in comprehensive modeling efforts on this important
research topic?
One approach to alleviate the apparent model--observational
discrepancy in coupled climate models is to improve mean-state
biases. The most overt form of this is to intervene in coupled GCMs and
flux correct them \citep[e.g.,][]{zhuo25}.
[In fact, \citet{knutson95} was a flux-corrected coupled GCM.]
Where do mean-state quantities appear in this
analysis? The climatological $SST$, $\mathcal{RH}$, and
$|\mathbf{u}|$ can influence the surface energy
budget response to perturbations.
Some biases, however, do not make straight-forward appearances. For
example, eliminating a double ITCZ or cloud bias may alter the radiation
component of the climatology of the meridionally averaged surface
energy budget, but does not constrain its changes---the
surface radiative feedback. Other pathways for a double ITCZ bias
to alter changes in the surface energy balance may
include changing the climatological surface wind or what would be higher-order
terms, like a different wind speed response to a radiative perturbation.


There have been proposals of missing or weak mechanisms in coupled climate models
to explain the apparent discrepancy in Pacific SST trends.
These include remote forcing [e.g., Southern Ocean glacial freshwater input
absent in models with unchanging Antarctic ice sheets \citep{dong22} or Southern Ocean
cooling teleconnections like those of ozone-induced changes in Southern Annual Mode \citep{hartmann22}],
which does not explicitly appear in the local surface energy balance.
For missing or weak mechanisms of influencing the tropical Pacific,
one can formulate these mechanisms as either a forcing or temperature-dependent change
on the tropical surface energy
budget, as was assessed for the Southern Ocean's influence on east Pacific in \citet{kang23}.
It is worthwhile to quantitatively state the magnitude of missing or weak mechanisms in these terms,
as it facilitates comparison with alternative mechanisms.
This is useful in the current situation, with multiple weak mechanisms, in contrast to
a situation where there are multiple quantitatively adequate mechanisms and all but one
must be ruled out.

Approximate surface energy balance calculations
provide quantitative pathways toward forced responses
of the climate system that are in line with observed trends
in the tropical Pacific SST.
The analysis here focused on a west-minus-east contrast
metric of Pacific SST trends, and there are additional
aspects of the trends, such as whether the east has cooling
rather than relatively weak warming, whose dynamics bear
further attention. 
The observed trends, of course, also have an internal variability component.
The pathways enumerated here should be subject to scrutiny
across observed changes in the atmospheric and oceanic
variables that underlie them. Clearly, they are also ripe for comparison
to comprehensive climate model simulations.
\acknowledgments I thank Nicole Feldl for connecting this research to
``storylines'' and Isaac Held and Allison Hogikyan for valuable discussions.
I am grateful to Matt Luongo and two anonymous reviewers for their thoughtful feedback.
I acknowledge support from the National Oceanic and
Atmospheric Administration, U.S. Department of Commerce under awards
NA18OAR4320123 and NA23OAR4320198.  The statements, findings,
conclusions, and recommendations are those of the author and do not
necessarily reflect the views of the National Oceanic and Atmospheric
Administration, or the U.S. Department of Commerce.

%
%
\datastatement
A repository of the code and input data to perform the surface energy balance calculations
and reproduce the figures is available at https://doi.org/10.5281/zenodo.15670801.

%






%



 \bibliographystyle{ametsocV6}
 \bibliography{references}

\end{document}